\documentclass{article}

\usepackage{a4}
\usepackage{epsf}
\usepackage{amsmath}
\usepackage{amssymb}

\newcommand{\PP}{{{\cal P}}}
\newcommand{\R}{{{\cal R}}}
\newcommand{\pt}{{p_t}}

\begin{document}

\title{Understanding of the freeze-out in ultra-relativistic %
heavy-ion collisions\thanks{Talk given at the 2nd Warsaw meeting on 
correlations and resonances, Oct. 15--18, 2003, Warsaw, Poland}}

\author{Boris Tom\'a\v sik\\[1ex]
{\it Niels Bohr Institute, Blegdamsvej 17,}\\
{\it DK--2100 Copenhagen \O , Denmark}}

\date{December 11, 2003}

\maketitle

\begin{abstract}
I discuss the quantities and effects important for the freeze-out
and outline a formalism for the description of continuous decoupling 
of particles from the fireball. Then I present a calculation of the 
scattering rates of pions at various temperatures and argue that 
it is important to take continuous particle decoupling into account
when modelling the collision dynamics.
\end{abstract}

Hadronic single-particle spectra and correlations 
carry information about the freeze-out state of the fireball which results
from the collective expansion 
of the strongly interacting matter. Often, the
freeze-out state is being modelled in framework of the 
Cooper-Frye \cite{Cooper:1974mv}
mechanism where all particles---regardless their identity 
and/or momentum---are emitted from a single sharp three-dimensional
freeze-out hypersurface. This is the case for most hydrodynamic simulations.
A question appears: does the assumption of a common sharp three-dimensional 
freeze-out still provide  reliable approximation of the real process?

In case of a sharp freeze-out, the freeze-out 
hypersurface is usually characterised by some prescription. There were
some attempts to identify a {\em universal freeze-out criterion}
\cite{Ferenc:1999ku,Adamova:2002ff}, i.e., a condition which
determines the freeze-out hypersurface for heavy-ion collisions
of any size and at any ultrarelativistic energy. 
In case of the continuous gradual decoupling there is no hypersurface 
to be determined and the concept of a universal freeze-out criterion 
is not applicable. I
will focus on gradual decoupling since the sharp freeze-out 
can be defined as its limiting case. 

Let me focus on the mechanism of freeze-out and 
identify the important effects and quantities. Freeze-out 
occurs when scattering ceases.
It has been suggested that the mean-free path is the relevant quantity 
to look at \cite{Adamova:2002ff,Nagamiya:tf}. Here, densities of the 
individual species are weighted with cross-sections for scattering
on them. An example: density of nucleons is much more important for pion 
scattering rate than the density of pions, because the $\pi$N cross-section
is bigger than the one for $\pi\pi$ scattering. The CERES collaboration 
argued that the ``universal'' pion mean-free path at freeze-out should be 
something {\em of the order of} 1~fm (maybe 2--3~fm) \cite{Adamova:2002ff}.

On the first sight, this is rather surprising, because this length is much 
shorter than the size of the system. So far, however, we did not mention 
the expansion and the decrease of the density due to it. In rough 
terms, freeze-out occurs when the rate of the density decrease becomes 
comparable or larger than the scattering rate \cite{Bondorf:kz}. 
The picture is the
following: at the moment of scattering the particle has some finite mean-free
path $\lambda$ and thus would be expected to scatter after some time. However,
if the density drops fast enough, after passing the length $\lambda$
our particle may find itself in an environment of very low density such that 
no other scattering occurs.

In \cite{Tomasik:2002qt} we used the formalism of escape probabilities
(conceived earlier by other authors, e.g. \cite{Grassi:1994nf,Sinyukov:2002if})
and focused on the dependence of the escape probability on temperature 
and the momentum of pions. By using this formalism one 
{\em qualitatively} goes beyond probing the mean-free path and the 
rate of density decrease; particles now decouple from a finite four-volume.
Sharp freeze-out is realised as a limiting case 
of this more general case, when the escape probabilities of {\em all} 
particles change from 0 to 1 in a very narrow space-time region.

The escape probability can be determined as 
\begin{equation}
\label{pdef}
\PP (x,\tau,p) = \exp \left ( - \int_\tau^\infty d\bar\tau\, 
\R(x+ v\bar \tau,p) \right )\, ,
\end{equation}
where $\R$ is the scattering rate. The opacity integral in this equation 
is evaluated along the expected trajectory of a particle with momentum 
$p$ and the position characterised by $x$ and $\tau$ if it moved straight. 
The integrated scattering rate gives
the average number of collisions the particle would suffer on this
trajectory. Assumptions about the chemical composition of the medium are
included in calculation of $\R$, while the density decrease rate determines 
its time dependence and thus the value of the resulting integral.

As an example of the fireball expansion we assumed in \cite{Tomasik:2002qt} 
a particle with $\pt = 0$ in the centre of the fireball which does
not move transversely but just waits until the surrounding matter decays.
Then we only need a prescription for the time dependence of $\R$ for
which we take an ansatz:
\begin{equation}
\R (\tau) = \R_0 \left ( \frac{\tau_0}{\tau} \right )^\alpha \, ,
\end{equation}
where $\alpha > 1$ and $\R_0$ is the scattering rate at time $\tau_0$. 
This corresponds 
to a fireball expanding longitudinally in a boost-invariant manner and 
transversely with the following dependence of the transverse 
flow rapidity on radial coordinate $r$:
\begin{equation}
\eta_t(r,\tau) = \frac{\alpha-1}{2}\frac{r}{\tau} \, .
\end{equation}
The escape probability in this toy model then reads
\begin{equation}
\label{expp}
\PP (\pt=0,r=0) = \exp \left ( - \frac{\R_0 \tau_0}{\alpha -1} \right ) =
\exp \left ( - \frac{\R_0}{2\eta_t(r,\tau_0)/r} \right )\, .
\end{equation}
Obviously, larger $\R_0$ leads 
to smaller escape probability, whereas stronger expansion is encoded
in larger $\alpha$ and/or $\eta_t$ and increases the value of $\PP$.

We calculated the scattering rate of negative pions in thermal hadronic
gas as
\begin{equation}
\label{req}
\R(p) = \sum_i \int d^3k\, \rho_i(k)\, \sigma_i'(s)\, |v_\pi - v_i|^*\, .
\end{equation}
Here, $\rho_i(k)$ is the density of particles on which the pion scatters,
$\sigma_i'(s)$ is the collinear cross-section, and $|v_\pi - v_i|^*$ is the
relative velocity of the pion to the other scattering partner in the CMS
of the pair. We integrate over the momenta of the scatterers and sum 
over following species as scattering partners: 
$i = \pi,\, N,\, \bar N,\, K,\, \rho,\, \Delta,\, \bar \Delta$.
For the cross-section we use a parametrisation of hadron scattering 
via resonances (see \cite{Tomasik:2002qt,Bass:1998ca}).
For the temperatures we assumed values of 90, 100, and 120~MeV. We estimated
the chemical potentials in such a way that we reproduced data on pion 
phase-space density at the SPS \cite{Murray&Holzer,wa98-psd} and 
RHIC \cite{lray} and the ratios of $dN/dy$ of different species at 
midrapidity \cite{Bachler:1999hu,Afanasiev:2002mx,Adcox:2001mf}.

\begin{figure}[t]
\epsfxsize=11cm
\centerline{\epsfbox{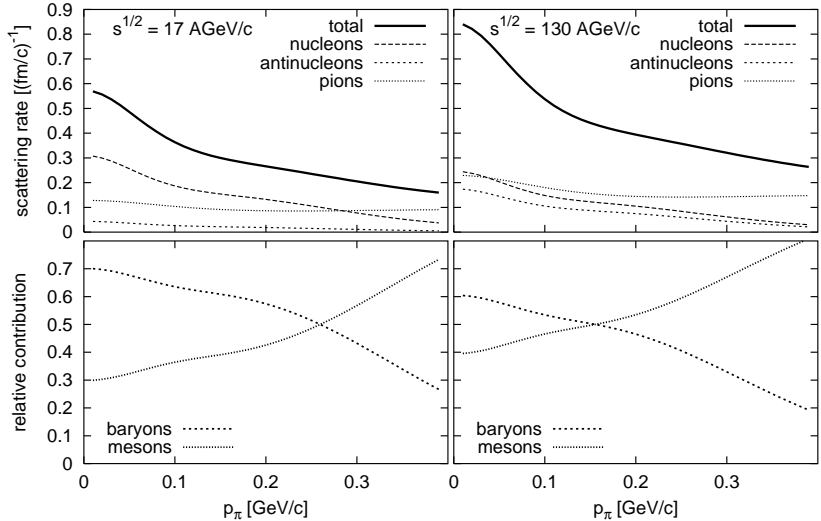}}
\caption{Scattering rate as a function of pion momentum 
with respect to the hadronic medium with $T=100$~MeV. Results are obtained
for SPS (left) and RHIC (right). Contributions to the scattering 
rate from scattering on pions, nucleons and antinucleons 
are indicated. The two lower panels show the baryonic and mesonic relative
contributions.
\label{fig1}}
\end{figure}
In Fig.~\ref{fig1} I compare pion scattering rates in hadronic
gases corresponding to those at the SPS and at RHIC. The  
contribution from nucleons and antinucleons to the total scattering
rate increases  little when moving from SPS to the higher energy system.
A higher phase-space density of pions at RHIC \cite{lray} leads to larger 
pion contribution, but the total scattering rate is not dominated 
by pions and therefore does not change much. Quantitatively, about 
10\% relative contribution is shifted from baryons to mesons when going 
from SPS to RHIC.

\begin{figure}[t]
\epsfxsize=7cm
\centerline{\epsfbox{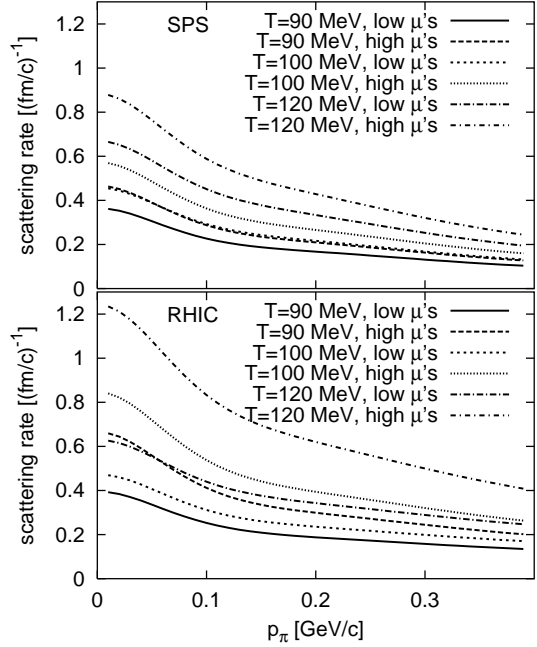}}
\caption{The pion scattering rates as functions of momentum
with respect to the medium, calculated for different temperatures
and sets of chemical potentials allowed by data. Details can 
be found in the original paper \protect\cite{Tomasik:2002qt}.
\label{fig2}}
\end{figure}
A summary of results is plotted in Fig.~\ref{fig2}.
The scattering rate {\em drops with temperature}. When we use
eq.~\eqref{expp} and a realistic estimate of the transverse rapidity 
gradient \cite{Tomasik:1999cq},  reasonable escape 
probabilities ($\sim $ 30--50\%) are obtained for $T \lesssim 100\,\mbox{MeV}$.
Note that this was inferred for a particle with $\pt=0$ in centre 
of the fireball.

One can see that regardless the assumed temperature and chemical potentials
higher momentum particles always have smaller scattering rate and thus 
decouple easier. This suggests that the freeze-out can be {\em sequential}:
high momenta first, low momenta later.
A sequential freeze-out might be an additional cause of the $M_t$ 
dependence of the HBT radii, if high-$\pt$ pions decouple earlier and from a
smaller fireball than the low-$\pt$ ones. Such a scenario, however,
must be studied in more detail. Nevertheless, it represents an interesting
alternative to the blast-wave model, which seems to have problems in 
reproducing data on HBT radii \cite{Tomasik:2003gt}. 

Due to the momentum dependence of the scattering rate, collapsing 
the whole decoupling four-volume into a sharp three-dimensional freeze-out
hypersurface seems an unreliable approximation.

\end{document}